\documentstyle[epsfig]{aipproc}

\newcommand{\be}{\begin{equation}}   \newcommand{\ee}{\end{equation}}
\newcommand{\bear}{\begin{eqnarray}}
\newcommand{\eear}{\end{eqnarray}}
\newcommand{\ba}{\begin{array}}      \newcommand{\ea}{\end{array}}

\newcommand{\CQ}{{\cal Q}}
\newcommand{\CU}{{\cal U}}
\newcommand{\CD}{{\cal D}}
\newcommand{\CL}{{\cal L}}
\newcommand{\CE}{{\cal E}}

\newcommand{\ov}{\overline}

\begin{document}

\vspace*{-1.2cm} \begin{flushright} EFI-2000-49 \end{flushright} 

\title{Electroweak Symmetry Breaking\\ and Extra Dimensions}

\author{Hsin-Chia Cheng}
\address{Enrico Fermi Institute, The University of Chicago,
Chicago, IL 60637}

\maketitle

\begin{abstract}
The electroweak symmetry may be broken by a composite Higgs which
arises naturally as a bound state of the top quark if the standard
model gauge fields and fermions propagate in extra dimenions.  The top
quark mass and the Higgs mass can be predicted from the infrared fixed
points of the renormalization group equations. The top quark mass is
in good agreement with the  experimental value, and the Higgs boson
mass is predicted to be $\sim 200$ GeV. The bounds on the
compactification scale can be quite low if all standard model fields
propagate in the same extra dimensions due to the momentum
conservation in extra dimensions. The current lower  limits are about
300 GeV for one extra dimension and 400-800 GeV for  two extra
dimensions. The future collider experiments may either discover the
Kaluza-Klein (KK) states of the standard model fields or raise their
mass limits significantly.  There may also be some other light bound
states which could be observed at upcoming collider experiments.
\end{abstract}

In the standard model (SM), the electroweak symmetry is broken by 
a nonzero vacuum expectation value of a fundamental scalar Higgs field. 
However, the squared mass of a scalar field receives quadratically
divergent radiative corrections, therefore suffers from the 
hierarchy problem if the cutoff is much higher than the weak scale.
Large extra dimensions accessible to the gravitons can remove the
large hierarchy between the weak scale and the Planck scale~\cite{ADD,RS}, 
but provide no understanding why the electroweak symmetry is broken,
{\it i.e.,} why there is a scalar field with the particular quantum
numbers and a negative squared mass. We will show that if the SM
fields also propagate in some extra compact dimensions of the 
size $\sim$ TeV$^{-1}$, a composite Higgs field can arise naturally 
as a bound state of the SM fermions. This is because gauge
interactions are non-renormalizable in more than four dimensions,
therefore rapidly become strong at energies not far above the 
compactification scale. The strong SM gauge interactions then can form
bound states from SM fermions. In particular, there is a Higgs bound state
composed of the top quark and its Kaluza-Klein (KK) excitations. This
provides the explanation of the Higgs quantum numbers~\cite{CDH,ACDH}.

Consider a one generation model in which the SM gauge fields
and the third generation fermions live in six dimensions, with two of
the six dimensions compactified at a scale $M_c \sim {\rm TeV}^{-1}$.
The theory is non-renormalizable hence needs a physical cutoff $M_s$. 
A possible candidate is the scale of quantum gravity, which is determined 
by the sizes of the extra dimensions accessible to the 
gravitons~\cite{ADD,RS}. 
In six dimensions, there exist four-component chiral fermions.
We assign $SU(2)_W$ doublets with positive chirality, $\CQ_+$,
$\CL_+$, and $SU(2)_W$ singlets with negative chirality, $\CU_-$,
$\CD_-$, $\CE_-$. Each fermion contains both left- and right-handed
two-component spinors when reduced to four dimensions. We impose an
orbifold projection such that the  right-handed components of $\CQ_+$,
$\CL_+$, and left-handed components of $\CU_-$, $\CD_-$, $\CE_-$, are
odd under the orbifold ${\bf Z}_2$ symmetry and therefore the
corresponding zero modes are projected out.  As a result, the
zero-mode fermions are two-component four-dimensional  quarks and
leptons: ${\cal Q}_+^{(0)} \equiv (t,b)_L$, \  ${\cal U}_-^{(0)}
\equiv t_R$, \  ${\cal D}_-^{(0)} \equiv b_R$, \  ${\cal L}_+^{(0)}
\equiv (\nu_\tau, \tau)_L$,  ${\cal E}_-^{(0)} \equiv \tau_R$.

At the cutoff scale, the SM gauge interactions are strong and will
produce bound states. The squared-mass of a scalar bound state has
quadratic dependence on the cutoff, and can become much smaller than
the cutoff scale or even negative if the coupling is sufficiently
strong.  Using the one-gauge-boson-exchange approximation, one finds
in general  among possible scalar bound states, that $H_{\CU}=\ov{\CQ}_+
\CU_-$, which has the correct quantum number to be the Higgs field, is
the most attractive channel. Therefore it is most likely to aquire a
negative squared-mass to break the electroweak symmetry. The composite
Higgs is expected to have a large coupling to its constituents, so the
theory not only predicts the correct Higgs quantum numbers, but also a 
heavy up-type quark (top quark). The $H_{\CD}= \ov{\CQ}_+ \CU_-$ channel 
is also quite strongly bound while the other channels are not
sufficiently strong to produce light bound states. The low-energy
theory below $M_c$ is expected to be a two-Higgs-doublet model.  In
more general models, there may be other light bound states  which may
be accessible at the upcoming collider experiments in addition to the
Higgs bosons.  The possible light bound states depend on the
model. For example, in the eight-dimensional model, there is a
strongly bound state  $\ov{\CQ} \CQ^c$ transforming like the
right-handed bottom quark under the SM gauge group~\cite{ACDH}.

Compared with the usual four-dimensional dynamical electroweak symmetry
breaking (EWSB)
models, the higher-dimensional model has the advantage that
the binding force can be the SM gauge interactions themselves,
without the need of introducing new strong interactions. In addition,
it also gives a prediction
of the top quark mass naturally in the right range. In the minimal
four-dimensional top quark condensate model, the top quark is too
heavy, $\sim 600$ GeV, 
if the compositeness scale is in the TeV range~\cite{BHL}.
With extra dimensions, the KK excitations of the top quark also
participate in the EWSB, so the top quark mass can be smaller.
Another way of understanding of the top Yukawa coupling being $\sim 1$
instead of the strong coupling value $\sim 4\pi$ is that (the zero
mode of) the top quark coupling receives a volume dilution factor
because it propagates in extra dimensions.  In fact, the top quark
mass can be predicted quite insensitively to the cutoff because of the
infrared fixed point behavior of the  renormalization group (RG)
evolution. The infrared fixed point is rapidly approached due to the
power-law running in extra-dimensional theories~\cite{DDG}, even
though the cutoff scale is not much higher than the weak
scale. Similarly, the Higgs self-coupling also receives  the
extra-dimensional volume suppression. As a result, the physical Higgs
boson is relatively light, $\sim$ 200 GeV, in contrast with the  usual
strongly coupled four-dimensional models. It is also governed by the
infrared fixed point of the RG equations. The numerical predictions of
the top quark mass and the Higgs boson mass are shown in Fig.~1.
\begin{figure}[htbp]
\centerline{\epsfysize=6.5cm\epsfbox{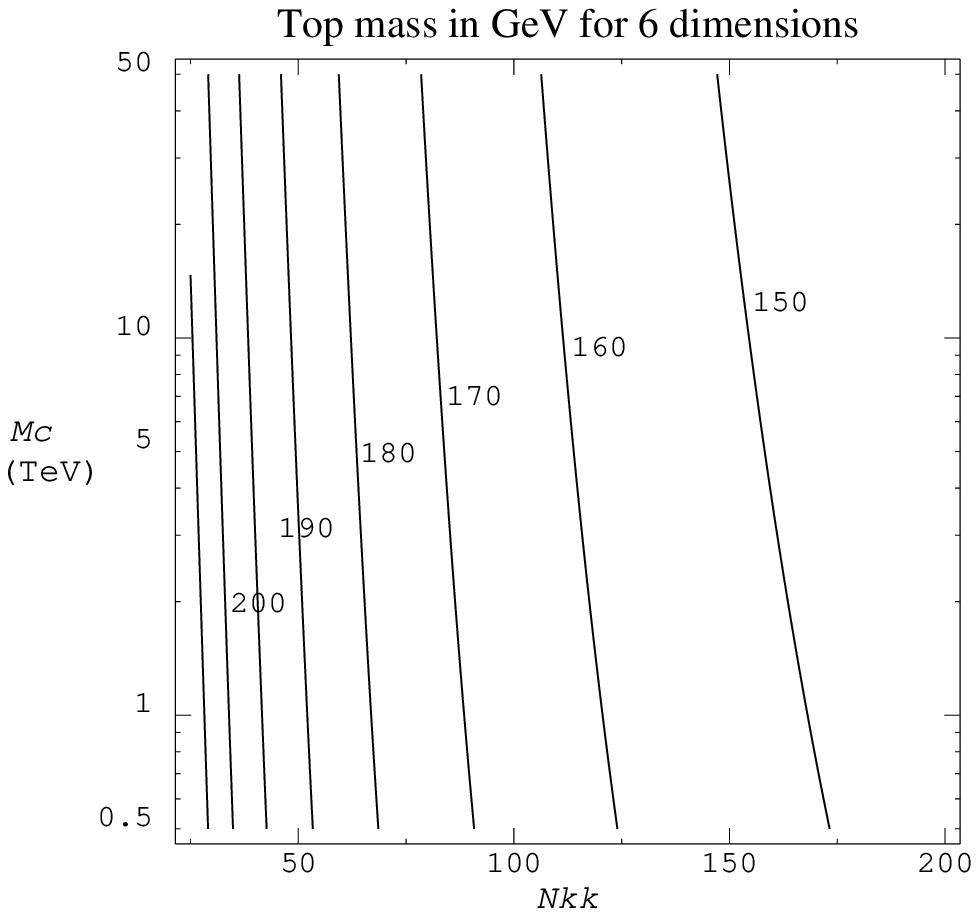}
\epsfysize=6.5cm\epsfbox{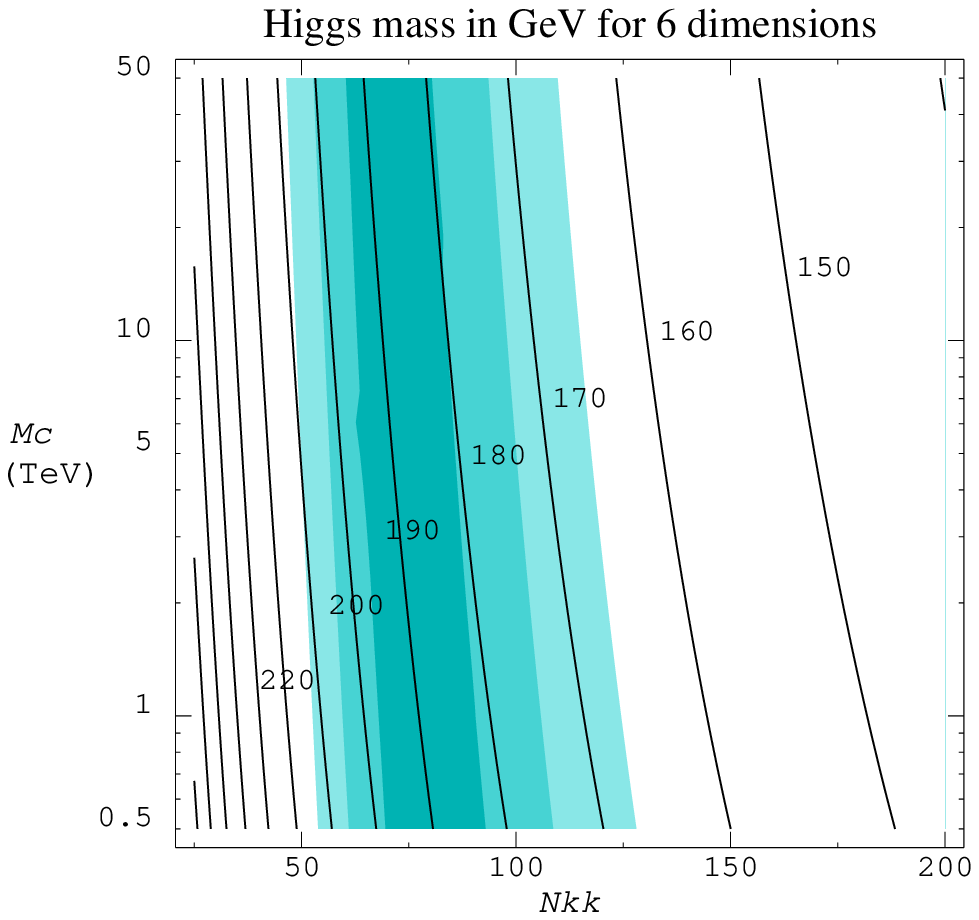}}
\caption{The top quark mass (left) and the Higgs boson mass (right)
as functions of the number of KK excitations, $N_{\rm KK}$, and the 
compactification scale $M_c$ in the six-dimensional theory. The shaded
area in the Higgs boson mass prediction corresponds to the top quark
mass lying within $3\sigma$ of the experimental value, $174.3\pm 5.1$~GeV.}
\end{figure}

So far we have not discussed the first two generation fermions.
If they are localized on some four-dimensional
subspaces, they can also form some four-dimensional 
bound states, though probably not as strongly bound as the higher
dimensional Higgs bound state, because they do not receive contributions
from the extra components of the gauge fields. Due to the tree-level
KK gauge boson exchange between the first two generation fermions,
there are strong bounds on the compactification scale, ${M_c \gtrsim}$ 
2--5 TeV from the precision electroweak data~\cite{bounds}.

Another interesting possibility is that all SM fields live in the
same extra dimensions (universal extra dimensions). In that case,  
we need to introduce explicit
flavor-breaking interactions from the cutoff scale to distinguish
the three generations. The flavor-breaking interactions should enhance
the third generation channels relative to the first two generation
channels so that only the composite Higgs field from the third generation
top quark have a negative squared-mass and is responsible for the
EWSB. Some flavor-breaking interactions
can also give masses to the other light fermions.

Because of the momentum conservation in extra dimensions, there is
no tree-level contribution to the electroweak observables from the
KK modes in the case of universal extra dimensions. There
are loop corrections because KK modes can appear in the loop. 
The experimental bounds on the size of the universal dimensions
are much weaker. The main constraints come from weak-isospin violation
effects. The lower limits for the compactification scales are
$\sim$ 300 GeV for one extra dimension and 400-800 GeV for two
extra dimensions~\cite{ACD}. With such a loose bound, there is a hope
that the upcoming collider experiments may discover the KK states.

At the colliders, the KK states have to be produced in pairs
or more because of the KK number conservation. If there is
no additional KK number violating interactions or the KK number violating
interactions are sufficiently weak, some of the KK modes will be stable
or long-lived. After being produced, they will hadronize into 
integer-charged states. The signatures will be highly ionizing
tracks. By extrapolating the current lower mass limits on heavy stable
quarks from the Run I of Tevatron~\cite{Connolly:1999dv}, we estimate
that the direct lower bound on the compactification scale is 300--350 GeV.
If there are KK number violating interactions which allow the KK modes
to decay inside the detector, the signatures and the limits depend
on the KK number violating effects. The direct bounds are somewhat
weaker, in the $\sim$ 200 GeV range~\cite{ACD}.
It is interesting that the direct limit is comparable to the indirect
bounds from the electroweak data, which means that the future collider
experiments may discover the KK modes or raise the lower bound 
significantly.

In summary, we have shown that if the standard model fields propagate
in extra dimensions of TeV$^{-1}$ size, the standard model gauge 
interactions become strong at high energies and naturally produce
a Higgs bound state from the top quark to break the electroweak
symmetry. The top quark mass and the Higgs boson mass can be predicted
from the RG infrared fixed points. The top quark mass is in good
agreement with the experimental value and the Higgs mass is around
200 GeV. There may be other light bound states which could be observed
in the future experiments. In the case all standard model
fields propagate in the same extra dimensions, the bounds on
the compactification scale are sufficiently loose that the KK modes
of the standard model fields may be discovered at the upcoming
experiments.




\end{document}